\documentclass[prd,aps,amsmath,amssymb,nofootinbib,twocolumn,preprintnumbers]
{revtex4}

\usepackage{graphicx}
\usepackage[lofdepth,lotdepth]{subfig}
\usepackage{dcolumn}
\usepackage{bm}
\usepackage{amsmath}
\usepackage{amsthm}
\usepackage{amsfonts}
\usepackage{euscript,bbm}
\usepackage{ifthen}
\usepackage{psfrag}
\usepackage{slashed}
\usepackage[normalem]{ulem}
\usepackage{lipsum}
\usepackage{hyperref}
\usepackage{multirow}
\usepackage{color}
\usepackage[utf8]{inputenc}
\usepackage{tikz,braket}
\usetikzlibrary{arrows,positioning, shapes.geometric}
\long\def\rpl#1!!#2!!{\textcolor{red}{#1} \textcolor{blue}{#2}}

\def \order(#1){{\cal O} \left(#1 \right)}

\allowdisplaybreaks

\begin{document}

\title{Higgs-Gluon Coupling in Warped Extra Dimensional Models with Brane Kinetic Terms}

\author{Ujjal Kumar Dey$^{a,}$\footnote{ujjaldey@prl.res.in}}
\author{Tirtha Sankar Ray$^{b,}$\footnote{tirthasankar.ray@gmail.com}}
\affiliation{$^{a}$Theoretical Physics Division, 
	          Physical Research Laboratory,\\
              Navrangpura, Ahmedabad 380009, India\\
             $^b$Department of Physics, 
		     Indian Institute of Technology, \\
		     Kharagpur 721302, India }

\begin{abstract}
	Warped models with the Higgs confined to the weak brane and  the gauge and matter fields accessing the $AdS_5$ bulk provide viable setting to address the   gauge hierarchy problem. Brane kinetic terms for the bulk fields are known to ease some of the tensions of these models with precision electroweak observables and flavor constraints. We study the loop-driven Higgs coupling to the gluons that are relevant to the Higgs program at LHC, in this scenario. We demonstrate a partial cancellation in the contribution of the fermionic Kaluza-Klein (KK) towers within such framework relatively independent of the 5D parameters. The entire dependence of this coupling  on the new physics arises from the mixing between the Standard Model states and the KK excitations. We find that the present precision in measurement of these couplings can lead to a constraint on the KK scale up to 1.2 TeV at $95\%$ confidence level.
\end{abstract}

\maketitle

\textbf{Introduction:}
Embedding the Standard Model (SM) in a slice of $AdS_5$ was originally introduced to utilize the non-factorizable 5D geometry in explaining the hierarchy between the bulk gravity scale and the SM scale confined in the weak brane \cite{Randall:1999ee}. However generalization of this scenario to allow the SM fermionic and gauge sectors to access the bulk have been extensively studied in the literature \cite{Gherghetta:2000qt, Huber:2000ie}. These models usually have the Higgs either bound to the TeV brane or identified with the fifth component of a bulk gauge boson to preserve the resolution of the gauge hierarchy problem. 
However, exotic states arising from integrating out the fifth spatial dimension usually contaminate the electroweak precision observable and the flavor structure of the SM. A non-trivial extension either of the bulk gauge group \cite{Agashe:2003zs} or the 5D space-time geometry \cite{Cabrer:2010si, Dillon:2014zea} is required to mellow these crippling constraints. A complementary approach to address some of these issues is to consider brane localized kinetic terms (BLKT)\cite{Fichet:2013ola, Carena:2004zn}.

The discovery of the Higgs \cite{Aad:2012tfa, Chatrchyan:2013lba} in the first phase of the LHC heralds the precision Higgs era. With the increasing statistics 
at the LHC and proposals for the construction of dedicated Higgs factories, the measured properties of the Higgs and its coupling to SM states will be at the center of the 
high energy physics program in the near future. Of crucial importance  is the coupling of the Higgs to the gluons as Higgs production through gluon fusion dominates at the LHC. Within the Standard Model they are
loop driven with third generation colored fermions dominating owing to their large Yukawa couplings. The Higgs-gluon vertex is in general vulnerable to the existence of new exotic 
colored states that have substantial coupling to the Higgs. The new states can show up as virtual particles in the loop and lead to considerable deviations from the SM 
expectation. The impact of an extra dimension on the couplings have  been  investigated in the literature \cite{Petriello:2002uu, Bhattacharyya:2009nb, Azatov:2010pf, Goertz:2011hj, Carena:2012fk, Malm:2014gha, Archer:2014jca}. 
Within the framework of the flat extra dimension it has been observed that inclusion of the BLKT results in a significant relaxation of the experimental constraints~\cite{Dey:2013cqa}.  

In this article we investigate the impact of the BLKT for the bulk fermions on the Higgs coupling to gluons within the warped extra dimensional framework. We find that the contribution of the Kaluza-Klein (KK) fermions to this coupling cancel. The  dependence of the coupling on the new physics scale entirely originates from the mixing between the SM states and the KK excitations. This contribution can be easily estimated analytically and imply  moderate constraints on the scale of new physics. This is in contrast with the scenario in the warped models without the BLKTs where extensions of the SM gauge group in the 5D bulk lead to large contribution to the gluon-Higgs couplings leading to severe constraints.

In the rest of the article we will briefly review the warped extra dimensional framework with BLKT. Then we will demonstrate the cancellation of the
direct contribution of the KK states to the Higgs-gluon coupling both analytically utilizing suitable approximations and also numerically using the exact relations. 
Finally we present the constraint on this framework by comparing the estimated coupling  with experimental results from the LHC,  before concluding.

\textbf{The model:} We consider a slice of $AdS_5$ where the 5D metric is given by, $ds^{2} = g_{MN}dx^{M}dx^{N} = e^{-2\sigma (y)}\eta_{\mu \nu}dx^{\mu}dx^{\nu}+dy^{2},$ where
$\eta_{\mu \nu} = {\rm diag}(-1,+1,+1,+1)$, $\sigma (y) = k|y|$ and $y\in [0,L]$.  The two 3-brane located at $y=0 (L)$ are the Planck (weak) branes respectively. The Higgs field is localized on  the weak  brane. 
The SM gauge bosons and fermions, along with gravity can access the bulk.  We assume that the bulk fermions have BLKTs. The action for the bulk  fermion is given by~\cite{Carena:2004zn},
\begin{align}
\mathcal{S}_{f} = &-i\int d^{4}x \int_{0}^{L} dy \sqrt{-g}\bigg[
                   \bar{\Psi}\Gamma^{A}e_{A}^{N}D_{N}\Psi 
                   +  M(y)\bar{\Psi}\Psi \notag\\
                  & + 2 \alpha_{f}
                     \delta(y-L)\bar{\Psi}_{L}\gamma^{a}
                     e_{a}^{\mu}\partial_{\mu}\Psi_{L}\bigg].
\end{align}
The determinant of the metric is denoted by $g$,  $\Gamma^{A}$ are the 5D  gamma matrices,  $A$ and $a$ represent the 5D and 4D Lorentz indices respectively, $e$ is the vierbien and $D_{N}$ is the covariant
derivative which contains the spin connection. The last term is the BLKT with coefficient $\alpha_{f}$ which has the dimension $[M^{-1}]$. 
Due to the requirement of the chirality of the fermion, the mass parameter $M(y)$ has to be an odd function of $y$ and is given by~\cite{Gherghetta:2000qt}, $M(y)=c_{f}\frac{d\sigma(y)}{dy}$ with $c_{f}$ being the parameter that determine the localization of the fermions in the bulk.
For simplicity we  introduce two  flavor independent BLKT coefficients into the theory:  (i) $\alpha_Q$ represents the common BLKT parameter for the $\mathrm{SU}(2)_L$ doublets, (ii) $\alpha_t$ represents the common BLKT parameters for the $\mathrm{SU}(2)_L$ singlet states. 
The Yukawa term is given by,
\begin{align}
\mathcal{S}_{\rm Yuk} = \int d^{4}x \int_{0}^{L} dy \sqrt{-g}
                         2\delta(y-L)y_{5}^i\bar{Q}^iH u^i,
                         \label{5dyukawa}
\end{align}
where $Q^i (u^i)$ are the 5D Dirac fields corresponding the left (right) chiral SM states of flavor $i$, we will suppress this index now to keep the discussion tractable.
 After KK decomposition, in the effective 4D theory these yield the usual massless chiral zero mode $q^{(0)}_L (u^{(0)}_R)$ and their
vector like KK excitations, $Q^{(n)}_L [+,+],~Q^{(n)}_R [-,-]  (U_R^{(n)} [+,+],~U_L^{(n)} [-,-])$ respectively. 5D perturbativity implies  $y_{5} \leq \sqrt{8}\pi \sqrt{\alpha_Q\alpha_t}/\sqrt{\log(\Lambda/k)}$, where  $\Lambda = e^{kL}\tilde{\Lambda}$ is $\mathcal{O}(M_{Pl})$ with $\tilde{\Lambda}$ being the cut-off. From the dimensional analysis it can be shown that $\Lambda/k \sim 25$~\cite{Carena:2003fx}. In our study we will assume a flavor diagonal Yukawa structure.
After integrating out the fifth spatial dimension one can obtain effective the 4D Yukawa coupling. Within the BLKT framework they can be obtained in closed form and were derived in \cite{Carena:2004zn}. Here we reproduce them for completeness,
\begin{align}
y^{i}_{q^{(0)}_{L} u^{(0)}_{R}} &= 
                 a_{Q_{L}}a_{u_{R}}\frac{y^{i}_{5} }{L} ,
                 ~~\mbox{where}~~\nonumber \\
a_{f} &= \sqrt{\frac{(1-2c_{f})kL \exp[(1-2c_{f})kL]}
         {\exp[(1-2c_{f})kL]\{1+(1-2c_{f})\alpha_{f}k\}-1}} \nonumber \\
y^{i}_{Q_{L}^{(n)}u_{R}^{(0)}} 
       &= a_{u_R}\frac{y^{i}_{5}}{L}\sqrt{\frac{2kL}
             {1+(1-2c^{i}_{Q_{L}})\alpha_{Q}k + 
              \alpha_{Q}^{2}k^{2}x^{i2}_{Q_{L}^{(n)}}}}, \nonumber \\
y^{i}_{Q_{L}^{(n)}U_{R}^{(m)}} 
       &= \frac{y^{i}_{5}}{L}\sqrt{\frac{2kL}
             {1+(1-2c^{i}_{Q_{L}})\alpha_{Q}k + 
              \alpha_{Q}^{2}k^{2}x^{i2}_{Q_{L}^{(n)}}}} \nonumber \\
           &  \times \sqrt{\frac{2kL}
             {1+(1-2c^{i}_{U_{R}})\alpha_{U}k + 
              \alpha_{U}^{2}k^{2}x^{i2}_{U_{R}^{(m)}}}},     
              \label{yukawas}         
\end{align}
where $i$ represents the flavor index and  $x^{i}_{Q_{L}^{(n)}/U_{R}^{(n)}}$ are the solutions of the transcendental equations that determine the mass of the $n$-th KK mode particle of flavor $i$ via the relation, $M^{i}_{f^{(n)}} = x^{i}_{f^{(n)}}\tilde{k}$ where $\tilde{k} = e^{-kL}k$. The transcendental equations are given by~\cite{Carena:2004zn},
\begin{subequations}
\begin{align}
J_{c^{i}_{f}-1/2}(x^{i}) &\approx \alpha_{f}kx^{i} J_{c^{i}_{f}+1/2}(x^{i})
      ~ \mbox{for} ~ c^{i}_{f} > \frac12,\\
J_{1/2-c^{i}_{f}}(x^{i}) &\approx -\alpha_{f}kx^{i} J_{-c^{i}_{f}-1/2}(x^{i})
      ~ \mbox{for} ~ c^{i}_{f} \leq \frac12.
\end{align}
\end{subequations}

\textbf{Gluon Fusion Cross Section:}
In the SM, the parton level cross section for the Higgs production via gluon fusion is given by~\cite{Gunion:1989we},
\begin{align}
\sigma_{gg\to H}^{\rm SM} = \frac{\alpha_{s}^{2}m_{H}^{2}}{576\pi}
                            \left|\sum_{i}\frac{y^{i}}{m^{i}}
                            A_{1/2}(\tau_{\rm SM}^{i})\right|^{2}
                            \delta(\hat{s}-m_{H}^{2}), 
\label{e:smhgg}
\end{align}
where the sum is over all SM fermions, $y^{i}$ and $m^{i}$ are the Yukawa coupling and mass of the SM fermion and  $\tau_{\rm SM}^{i} = \left(\frac{m_{H}}{2m^{i}}\right)^{2}$ and the function $A_{1/2}$ is defined as,
\begin{align}
A_{1/2}(\tau) &= \frac32 \tau^{-2}[\tau+(\tau -1)f(\tau)], \nonumber \\~~\mbox{where}~~
f(\tau) &= \begin{cases}
           \left(\sin^{-1}\sqrt{\tau}\right)^{2}~\mbox{for}~\tau\leq1,\\
           -\frac14\left[\ln\left(\frac{1+\sqrt{1+\tau^{-1}}}
              {1-\sqrt{1-\tau^{-1}}}\right)
              -i\pi\right]^{2}~\mbox{for}~\tau > 1.
          \end{cases}
    \label{ahalf}
\end{align}

In the warped extra dimensional scenario there are three main sources of deviation in the Higgs coupling to gluons: 
(i) the  KK fermions obtained by integrating out the fifth dimension  will contribute in the loop,  (ii)  modified contribution 
from the zero modes due to the fact $y_{\rm RS}^{i} \neq m^{i}/v_{\rm SM}$ due to mixing in the Yukawa sector and (iii) the modification 
in the Higgs vev due to mixing in the gauge sector $\tilde{v} \neq v_{\rm SM}.$ Note that the KK masses  $m^{(n)}\gg m_{H}$, 
implying  $A_{1/2} \rightarrow 1$. Incorporating these factors  requires the following modification of  Eq.~\ref{e:smhgg}, 
\begin{align}
\frac{y^{i}}{m^{i}} A_{1/2}(\tau_{\rm SM}^{i})
                     &\to \frac{y^i_{\rm RS}}{m^i_{\rm RS}}
                        A_{1/2}(\tau_{\rm SM}^{i})
                        +\sum_{\rm KK}\frac{Y^{i}}{M^{i}}\nonumber \\
                     &=  {\rm Tr}\left(\tilde{Y^i}\tilde{M^i}^{-1}\right)
                       +\frac{y^i_{\rm RS}}{m^i_{\rm RS}}
                        \left(A_{1/2}(\tau^i_{\rm SM})-1\right),
                        \label{totalKK}
\end{align}
where $\tilde{Y}$ and $\tilde{M}$ are the Yukawa and fermion mass matrices and $Y$ and $M$ are their eigenvalues. Note that the piece $\left(\text{Tr}(\tilde{Y}^{i}/\tilde{M}^{i})-y^{i}_{\rm RS}/m^{i}_{\rm RS}\right)$ encodes the effect from the loop exchange of the heavy eigenstates barring the zero mode. The mass and the Yukawa matrices for the quark sector are given by,

\begin{widetext}
\begin{align}
\label{massmatrix}
\tilde{M}^i = \begin{pmatrix}
\frac{1}{\sqrt{2}}y^i_{q_{L}^{(0)}u_{R}^{(0)}}\tilde{v}  &  0  &  
\frac{1}{\sqrt{2}}y^i_{q_{L}^{(0)}U_{R}^{(1)}}\tilde{v}  &  0  & 
\frac{1}{\sqrt{2}}y^i_{q_{L}^{(0)}U_{R}^{(2)}}\tilde{v}  & \ldots \\
\frac{1}{\sqrt{2}}y^i_{Q_{L}^{(1)}u_{R}^{(0)}}\tilde{v}  &  M^i_{Q^{(1)}}  &
\frac{1}{\sqrt{2}}y^i_{Q_{L}^{(1)}U_{R}^{(1)}}\tilde{v}  &  0  & 
\frac{1}{\sqrt{2}}y^i_{Q_{L}^{(1)}U_{R}^{(2)}}\tilde{v}  & \ldots  \\
0  &  \frac{1}{\sqrt{2}}y^i_{U_{L}^{(1)}Q_{R}^{(1)}}\tilde{v}  &  M^i_{U^{(1)}}  &
\frac{1}{\sqrt{2}}y^i_{U_{L}^{(1)}Q_{R}^{(2)}}\tilde{v}  &  0  & \ldots \\
\frac{1}{\sqrt{2}}y^i_{Q_{L}^{(2)}u_{R}^{(0)}}\tilde{v}  &  0  &
\frac{1}{\sqrt{2}}y^i_{Q_{L}^{(2)}U_{R}^{(1)}}\tilde{v}  &  M^i_{Q^{(2)}}  & 
\frac{1}{\sqrt{2}}y^i_{Q_{L}^{(2)}U_{R}^{(2)}}\tilde{v}  & \ldots  \\
  &  &  \vdots  &  & \ddots
\end{pmatrix},
\end{align}
\end{widetext}

and 
$\tilde{Y}^i = \frac{\partial \tilde{M}^i}{\partial \tilde{v}},$ where $q_L^{(0)}~\mbox{and}~ u_R^{(0)}~(Q_L^{(n)} ~\mbox{and}~ U_R^{(n)})$ represent the SM (KK excitations)  left chiral $\mathrm{SU}(2)_{L}$ doublet and  right handed singlet states respectively. We neglect flavor mixing in our study. The modified values of the SM parameters due to mixing in the Yukawa sector is given by, 
\begin{align}
\label{e:mRS}
m_{\rm RS}^i = \frac{1}
                {\sqrt{2}}y^i_{q_{L}^{(0)}u_{R}^{(0)}}\tilde{v}
                 &+ \sum_{m,n}\bigg[y^i_{q_{L}^{(0)}U_{R}^{(n)}}
                   \frac{1}{M^{i}_{U_{R}^{(n)}}}
                     y^i_{U_{R}^{(n)}Q^{(m)}}
                   \nonumber \\
                 &\times \frac{1}{M^{i}_{Q_{L}^{(m)}}}
                 y^i_{Q_{L}^{(m)}u_{R}^{(0)}}
                   \left(\frac{\tilde{v}}{\sqrt{2}}\right)^{3}
                 \bigg]
\end{align} 
 and $
y^i_{\rm RS} = \frac{\partial m_{\rm RS}^i}
                      {\partial \tilde{v}}. $
The quantity $\tilde{v}$ is the modified Higgs vev arising from the mixing with KK states in the gauge sector. It can be approximately given by~\cite{Azatov:2010pf},
\begin{align}
\tilde{v} \approx v_{\rm SM}\left(1 + 
              \frac{(g_{5}\sqrt{k})^{2}v_{\rm SM}^{2}}
              {16\tilde{k}^{2}}\right),
\label{vtilde}
\end{align}
where  $g_{5}\sqrt{k} \sim 6$~\cite{Gherghetta:2010cj} is set from the measured masses of the SM gauge bosons.
Considering only up to first KK mode which is expected to constitute the leading contribution of the extra dimension, due to the efficient decoupling of the higher KK modes within the BLKT framework \cite{Carena:2004zn},
 the term ${\rm Tr}\left(\tilde{Y}^i(\tilde{M}^i)^{-1}\right)$ 
can be given as (keeping up to $\mathcal{O}(v^{2}_{\rm SM}/ \tilde{k}^{2})$)~\cite{Azatov:2010pf},
\begin{widetext}
\begin{align}
\label{e:1leveltr}
{\rm Tr}\left(\tilde{Y}^i(\tilde{M}^i)^{-1}\right) &\approx 
               \frac{1}{\tilde{v}}\left[ 1+
                \left(\frac{y^i_{Q_{L}^{(1)}u_{R}^{(0)}}
                 y^i_{U_{L}^{(1)}Q_{R}^{(1)}}
                  y^i_{q_{L}^{(0)}U_{R}^{(1)}}}
                   {y^i_{q_{L}^{(0)}u_{R}^{(0)}}}
                    - y^i_{Q_{L}^{(1)}U_{R}^{(1)}}
                       y^i_{U_{L}^{(1)}Q_{R}^{(1)}}\right)
               \frac{\tilde{v}^{2}}{M^i_{Q_{L}^{(1)}}M^i_{U_{R}^{(1)}}} \right]
               \notag \\
               &= \frac{1}{\tilde{v}} \left[
                    1 + (\mathcal{X}^i-\mathcal{X}^i)
                    \frac{\tilde{v}^{2}}
                    {\tilde{k}^2}\right] = 
                    \frac{1}{\tilde{v}} \approx 
                    \frac{1}{v_{\rm SM}}\left(
                    1-\frac{9}{4}\frac{v^{2}_{\rm SM}}
                    {\tilde{k}^{2}}\right),\\
\mbox{where}\nonumber\\ 
\label{xi}
\mathcal{X}^i &=\frac{1}{{x^i_{Q_{L}^{(1)}}x^i_{U_{R}^{(1)}}}}
                 \frac{4y_{5}^{i2}k^{2}}
               {\left[1 + (1-2c^i_{Q_{L}})\alpha_{Q}k
                 + \alpha_{Q}^{2}k^{2}x_{Q_{L}^{(1)}}^{i2}
                  \right]\left[1 + (1-2c^i_{U_{R}})\alpha_{U}k
                 + \alpha_{U}^{2}k^{2}x_{U_{R}^{(1)}}^{i2}
                  \right]}.
\end{align}
\end{widetext}
This expression of $\mathcal{X}^i$ can be derived using Eqs.~\ref{yukawas}. 
Now, following Eq.~\ref{e:mRS}, one can write, within the same  approximation, an expression for $y_{\rm RS}^{i}/m_{\rm RS}^{i}$ in terms of $\mathcal{X}^{i}$, given by,
\begin{align}
\frac{y_{\rm RS}^{i}}{m_{\rm RS}^{i}} \approx \frac{1}{\tilde{v}}
                     \left[1+\mathcal{X}^{i}
                     \frac{\tilde{v}^{2}}{\tilde{k}^{2}}\right]
                     \approx \frac{1}{v_{\rm SM}}\left[
                     1+\left(\mathcal{X}^{i}-\frac{9}{4}\right)
                     \frac{v^{2}_{\rm SM}}{\tilde{k}^{2}}\right].
                     \label{mby}
\end{align}
From Eq.~\ref{e:1leveltr} we see that $\text{Tr}(\tilde{Y}^{i}/\tilde{M}^{i}) = 1/\tilde{v}$, which implies that the contribution from the Yukawa sector of the KK fermion vanishes at the order $v_{\rm SM}^{2}/\tilde{k}^{2}$ and this is independent of the flavor index and choice of the BLKT parameters. However, the $Hgg$ coupling still depends on the detail of the KK spectrum which enters the amplitude expression through Eq.~\ref{mby}, thus rendering the cancellation to be partial. 
This conclusion remains unchanged once we include the full KK tower and constitutes the main observation made in this article. 
We point out that the cancellation, alluded to in Eq.~\ref{e:1leveltr}, is more general than the BLKT framework that is being discussed here. 
It  is a characteristic of models with vector-like top partners, where there is a  cancellation between the effect from the  mixing with top quark and the direct contribution in loops, for a detailed discussion see~\cite{Aguilar-Saavedra:2013qpa}. However, the cancellation in  Eq.~\ref{e:1leveltr}, within the present scenario  of a brane Higgs with 
the Yukawa Lagrangian given in Eq.~\ref{5dyukawa} is a result of the generic structure of the Yukawa couplings  given by
$y_{Q^{(n)}U^{(m)}} = N f_{Q^{(n)}}(c_Q,\alpha,L)f_{U^{(m)}}(c_U,\alpha,L),$ where $ f_{(Q/U)^{(n)}}(c,\alpha,y)$ is the bulk profile of the 
$(Q/U)^{(n)}-$th state. In models without BLKT, usually the bulk gauge groups is augmented to control the electroweak observables. This necessitates the introduction of additional custodial partners and their KK towers. The contributions from these new states do not cancel, making the result dependent on the details of the theory.
In addition to the cancellation of the KK mode contributions, the effect of mixing in the top  Yukawa sector  effectively cancels due to the fact that  $A_{1/2} \sim 1$ in Eq.~\ref{totalKK},  remains a  good approximation for the top quark. 
We are left with the  modification of the Higgs vev, Eq.~\ref{vtilde} due to mixing in the gauge sector. The other remaining deviation from SM arises due to  mixing in the Yukawa sector for the light quarks, as obtained by substituting  Eq.~\ref{mby} in Eq.~\ref{totalKK}. We find the ratio of the cross section in the RS scenario to the SM value $C_{gg} = \sigma^{\rm RS-BLKT}(gg\rightarrow H)/\sigma^{\rm SM}(gg\rightarrow H)$ is approximately given by (considering the contribution form the 1st KK level only),
\begin{align}
C_{gg} \sim \bigg|1- \left(\frac{v_{\rm SM}}{\tilde{k}} \right)^2 &\left(\frac{9}{4}+\sum_{\rm light-fermion} {\mathcal{X}}^i\right) 
\nonumber \\ &+{\mathcal{O}}\left(\left(\frac{v_{\rm SM}}{\tilde{k}} \right)^4\right)\bigg|^2,
\label{cgg1}
\end{align}
where by the sum over light fermions we mean the sum over all quarks except top and we have assumed $A_{1/2} \sim 0$ for the light quarks. The new physics scale  is defined as $\tilde{k} =k \exp[-\pi kR].$
It is straightforward to generalize this expression in Eq.~\ref{cgg1} to include the contribution of the higher  KK modes. This requires the following modification of the above equation,
\begin{equation*}
\label{eq:X}
{\mathcal{X}}^i\to \sum_{m,n}\frac{y^i_{q^{(0)}U^{(n)}}y^i_{U^{(n)}Q^{(m)}}y^i_{Q^{(m)}u^{(0)}}}{y^i_{q^{(0)}u^{(0)}}x^i_{Q^{(m)}}x^i_{U^{(n)}}}
\end{equation*} 
The leading effect of the  BLKT  on the Higgs-gluon coupling is parametrized by the factor $\mathcal{X}^i$ in the above expressions. However, due to the structure of the theory, the contribution from the light fermions is small compared to the contribution 
from mixing in the gauge sector. Thus a very good approximation is   
\begin{equation}
\label{eq:cggapp}
C_{gg}^{app} \sim 1- \frac{9}{2} \frac{v_{\rm SM}^2}{\tilde{k}^2},
\end{equation}
which is fairly independent of the details of the 5D theory.

\begin{figure*}[!htbp]
\subfloat[Subfigure 1 list of figures text][]{
\includegraphics[width=7.0cm,height=6.0cm]{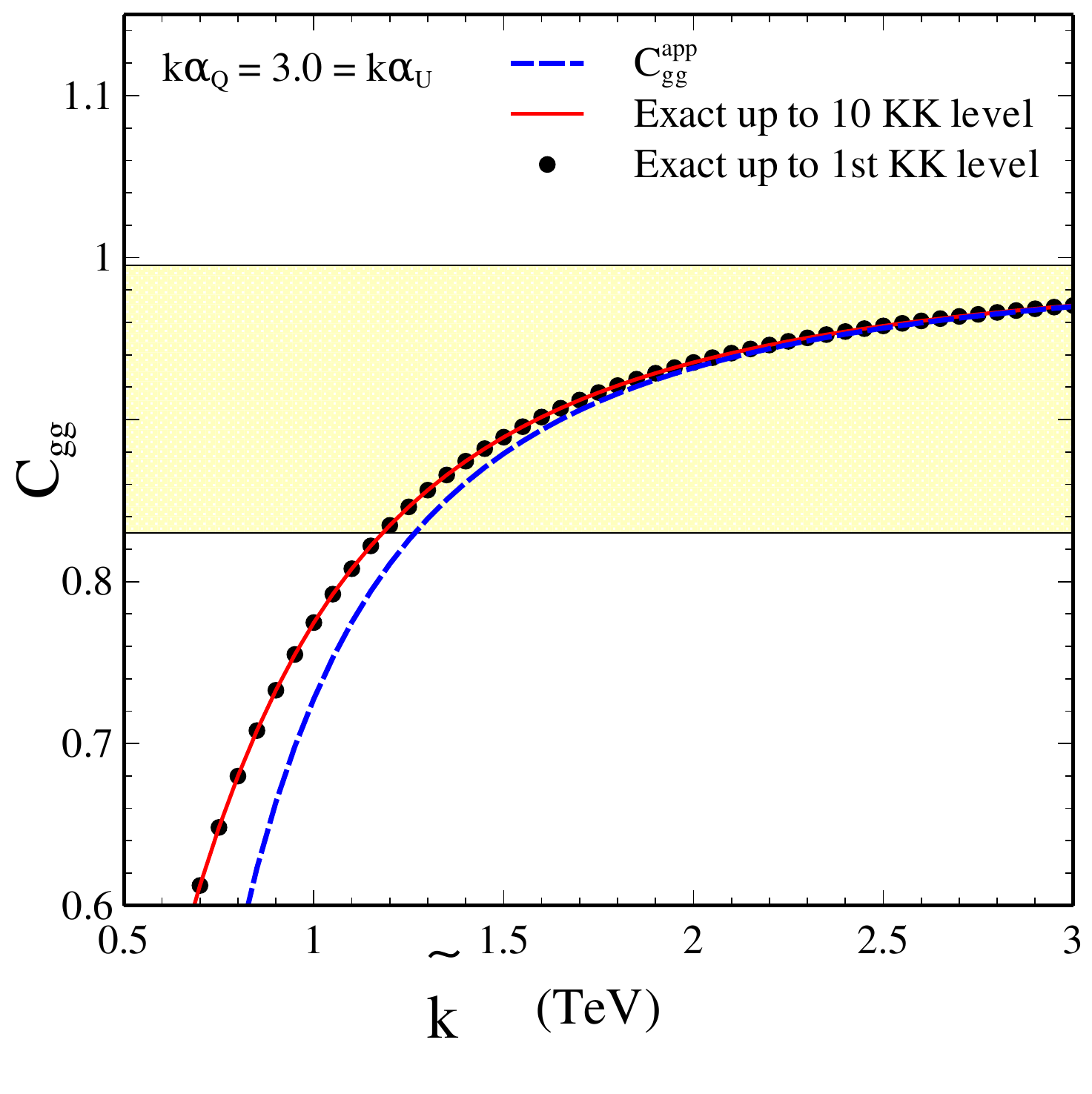}
\label{fig1a}}
~~~~
\subfloat[Subfigure 2 list of figures text][]{
\includegraphics[width=7.0cm,height=6.1cm]{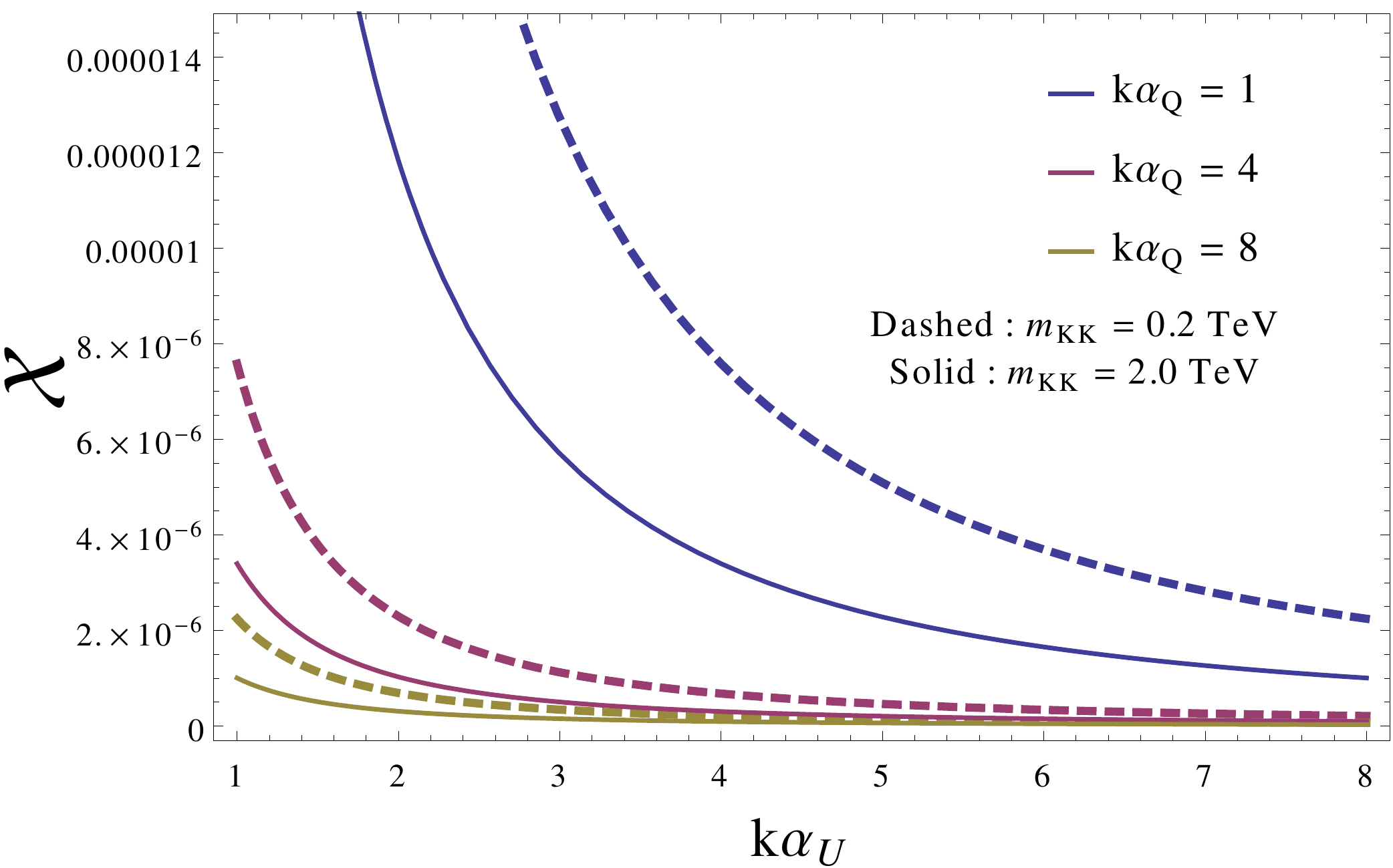}
\label{fig1b}}
\caption{\small In the left panel (a) we plot the  $gg\rightarrow h$ cross section in the  warped model with BLKT for bulk fermions normalized by the SM value against the new physics scale $\tilde{k} .$ The shaded band corresponds to the 2$\sigma$ bound on $C_{gg}$ obtained from \cite{Ellis:2013lra}, also see \cite{Carmi:2012in}. The solid red line and the black dots are obtained from the full numerical simulation keeping ten and one KK modes respectively. The blue (dashed) line corresponds to the approximation given in Eq.~\ref{eq:cggapp}. The right panel (b) shows the parameter $\mathcal{X} =\sum_{i}\mathcal{X}^{i}$ plotted as a function of the BLKT parameters for two different values of $\tilde{k}$. This corresponds to the leading contribution of the BLKT parameters to the Higgs-gluon coupling and should be numerically compared with the contribution from mixing in the gauge sector $\sim 9/2$.}
\label{fig:globfig}
\end{figure*}

\textbf{Results:}
We perform a detailed numerical calculation of the new physics (NP) contribution of the Higgs-gluon coupling by numerically diagonalizing the full mass matrix as given in Eq.~\ref{massmatrix}. Clearly the numerical result includes all orders of $v^{2}_{SM}/\tilde{k}^{2}$. 
We assume, for simplicity a left right symmetry for the 5D localization of the bulk fermions, {\it i.e.}, $c_Q^i = c_U^i .$ 
We fix the values  of these by comparing with the measured masses of the SM fermion at the limit where the BLKT parameters vanish and the 5D Yukawa couplings is universally set to unity  \cite{Ray:2011sz}, to reproduce all the SM quark masses. The  numerical values of these $c$-parameters are given in Table~\ref{ctable}.
\begin{table}[!htbp]
\centering
\begin{tabular}{|c|c|c|c|c|c|c|}
\hline
$f^{i}$ & $u$  & $d$  & $c$  & $s$  & $t$   & $b$  \\ \hline \hline
$c^{i}$ & 0.62 & 0.57 & 0.52 & 0.52 & -0.50 & 0.26 \\ \hline \hline
\end{tabular}
\caption{The $c$-parameters for various quarks~\cite{Ray:2011sz}.}
\label{ctable}
\end{table}
Once the BLKT parameters are turned on the eigenvalues of the mass matrix shift. The resultant deviation from the SM masses  are compensated by fixing the 5D Yukawa coupling of every flavor in  Eq.~\ref{5dyukawa}, for a given value of the BLKT parameters and the NP scale. We also check that the normalized 5D Yukawas stay within the perturbative limits. We numerically calculate the KK contributions from every flavor, including the light quarks which have large KK Yukawa couplings. 
In our numerical analysis we  keep the first 10 KK modes in the mass matrix. We compare our results with the approximate analytical expression derived in the previous section. The results of the full numerical simulation and the approximate theoretical results are plotted in Fig.~\ref{fig1a}. We find that the constraint on the
new physics scale $\tilde{k}$ is given at $\sim 1.2$ TeV. 
In Fig.~\ref{fig1b} we demonstrate  the impact of the BLKT parameters, parametrized by $\mathcal{X}=\sum_{i}\mathcal{X}^{i}$. We find that the residual impact of the BLKT parameter on the Higgs-gluon
coupling is negligible compared to contribution of the mixing in the gauge sector {\it i.e.}, $\mathcal{X} \ll 9/2$. This allows us to choose the BLKT parameters  for the bulk fermions from the precision electroweak fit, independent of the prediction for the Higgs-gluon coupling. The choice of $k\alpha_Q, k\alpha_U > 3$ results in an efficient suppression of the KK fermion contribution to the oblique electroweak precision observables \cite{Carena:2004zn}. For example the contribution to the $T$ parameter from the first KK  top mode in the loop is given by $\Delta T \sim (y_{Q_L^{(1)}U_R^{(1)}}/y_t)^4(m_t/\tilde{k})^2{\mathcal{O}}(1) \sim 1.1\times10^{-6}~(1\times10^{-8})$ for BLKT parameters  $3~(10)$ and $\tilde{k}=1$ TeV. Similarly the contribution due to the mixing between the first KK mode and the zero mode in the top sector is schematically given by $\Delta T \sim (y_{Q_L^{(1)}t_R^{(0)}}/y_t)^2(m_t/\tilde{k})^2{\mathcal{O}}(10) \sim 0.01 ~(0.003)$ respectively. This easily evades the present bound $\Delta T < 0.13$~\cite{Baak:2014ora}. However, note that there could still be potentially a large contribution from the gauge sector depending on the details, including the choice of the corresponding BLKT parameters.

\textbf{Conclusion:} The Higgs coupling to gluons is crucial for the Higgs physics program at the LHC. The impact of new exotic colored states in various new physics scenarios can lead to a significant modification of the Higgs couplings. The ever decreasing experimental uncertainty on these couplings will lead to either a non-trivial indication for new states or constrain them. In this context we revisit the Higgs coupling in the warped extra dimensional scenario. We assume that the Higgs is localized on the weak brane and the other SM states can access the bulk. The SM fermions have profiles that are localized in the bulk providing a handle on the fermion mass hierarchy issue. The bulk fermions have brane localized kinetic terms that can restrain the constraints from the electroweak precision observables. We show that the contribution from the KK states on the loop-driven Higgs coupling to the gluon vanishes. The residual dependence on the extra dimensional parameters arise from the mixing of the SM fermionic and gauge states with their KK excitations. The mixing in the gauge sector dominates and lead to a modification that  is proportional to ${\mathcal{O}}(1)v^2/\tilde{k}^2,$ where $\tilde{k}$ is the scale of new physics.  We find that the order one coefficient is rather independent of the details of the  5D theory. We have performed an extensive numerical calculation of this Higgs coupling. We find that the present experimental bounds at 2$\sigma$ confidence disfavor a new physics scale below $\sim 1.2$ TeV which is relatively independent of choice of the extra dimensional parameters like the bulk mass terms and the BLKT coefficient for the fermions. However, in this article we have not considered the introduction of BLKT for the gauge sector or brane mass terms that usually split the KK spectrum. These may lead to non-trivial modification of the loop-driven Higgs couplings, a careful study of which is now warranted. It should also be noted that the Higgs field in the warped framework is also modified due to mixing with the radion \cite{deSandes:2011zs, Desai:2013pga, Cox:2013rva}, which we neglect in the present analysis.
In the end we point out the curious similarity between the cancellation and the residual corrections obtained in this scenario with  the Higgs couplings in the gauge Higgs unification scenario \cite{Falkowski:2007hz}, which are dual to the composite Higgs framework where the Higgs is identified as a pseudo-Nambu-Goldstone mode of a strong sector \cite{Montull:2013mla, Azatov:2011qy}.

\paragraph*{\bf Acknowledgements\,:} We thank Tony Gherghetta and Alexandr Azatov for discussions. TSR thanks ICTP, Trieste, Italy for hospitality during the completion of this work. The work of TSR is supported by an ISIRD grant IIT-Kharagpur, India.


\bibliographystyle{h-physrev}
\bibliography{ref}

\end{document}